\documentclass[aapm,mph,preprint,graphicx,nofootinbib]{revtex4-1}
\usepackage[mathlines]{lineno}
\usepackage{epsfig}

\begin{document}
\title{A fast GPU-based Monte Carlo simulation of proton transport with detailed modeling of non-elastic interactions}

\author{H.~Wan Chan Tseung}
\email{wanchantseung.hok@mayo.edu}
\author{J.~Ma}
\author{C.~Beltran}
\affiliation{Department of Radiation Oncology, Mayo Clinic, Rochester MN 55905}
\date{\today}

\begin{abstract}
\noindent {\bf Purpose:} Very fast Monte Carlo (MC) simulations of proton transport have been implemented recently on graphics processing units (GPUs). However, these MCs usually use simplified models for non-elastic proton-nucleus interactions. Our primary goal is to build a GPU-based proton transport MC with detailed modeling of elastic and non-elastic proton-nucleus collisions.
\\\noindent {\bf Methods:} Using the CUDA framework, we implemented GPU kernels for the following tasks: (1) Simulation of beam spots from our possible scanning nozzle configurations, (2) Proton propagation through CT geometry, taking into account nuclear elastic scattering, multiple scattering, and energy loss straggling, (3) Modeling of the intranuclear cascade stage of non-elastic interactions when they occur, (4) Simulation of nuclear evaporation, and (5) Statistical error estimates on the dose. To validate our MC, we performed: (1) Secondary particle yield calculations in proton collisions with therapeutically-relevant nuclei, (2) Dose calculations in homogeneous phantoms, (3) Re-calculations of complex head and neck treatment plans from a commercially-available treatment planning system, and compared with Geant4.9.6p2/TOPAS. 
\\{\bf Results:} Yields, energy and angular distributions of secondaries from non-elastic collisions on various nuclei are in good agreement with the Geant4.9.6p2 Bertini and Binary cascade models. The 3D-gamma pass rate at 2\%-2 mm for treatment plan simulations is typically 98\%. The net computational time on a NVIDIA GTX680 card, including all CPU-GPU data transfers, is $\sim$20 s for $1\times10^7$ proton histories.
\\{\bf Conclusions:} Our GPU-based MC is the first of its kind to include a detailed nuclear model to handle non-elastic interactions of protons with any nucleus. Dosimetric calculations are in very good agreement with Geant4.9.6p2/TOPAS. Our MC is being integrated into a framework to perform fast routine clinical QA of pencil-beam based treatment plans, and is being used as the dose calculation engine in a clinically-applicable MC-based IMPT treatment planning system. The detailed nuclear modeling will allow us to perform very fast linear energy transfer and neutron dose estimates on the GPU.
\end{abstract}

\maketitle


\section{Introduction}
The ability to perform fast and accurate dose calculations is of great importance to modern radiation therapy, especially in proton therapy, where range uncertainties can be critical. In terms of accuracy, it is widely accepted that Monte Carlo (MC) techniques are the gold standard for dose calculations. MC simulations of proton treatment plans have previously been performed using well-established packages such as Geant4, FLUKA and MCNPX \cite{Paganetti2008, fluka, MCNP}. The improved accuracy of MC simulations compared to pencil beam calculations has been explicitly demonstrated \cite{Paganetti2008}. Because of the enhanced accuracy, MC dose calculations can be used to verify results from pencil beam-based treatment planning systems (TPS), hence mitigating the effects of proton range uncertainties \cite{Paganetti2012}. Unfortunately, computational times associated with MC-based treatment plan simulations using Geant4, FLUKA and MCNPX can be prohibitive.  For example, a TOPAS \cite{TOPAS} simulation of the dose within 2\% statistical error ($\sim$1$\times10^7$ particle histories) in a 150 cm$^3$ tumor volume typically requires around 1 hour on a 100-node CPU cluster. This makes routine clinical QA of treatment plans very difficult. At best, MC evaluations can be reserved for a handful of cases, and even so, only if significant computing resources are available.

Recently, several attempts to implement condensed history MC simulations of proton transport on graphics processing units (GPUs) have been made, with very encouraging results \cite{Kohno, Jia, Su, Osiecki}. Using Berger's classification, these attempts can be grouped as class I \cite{Kohno, Su, Osiecki} and II \cite{Jia}. The typical processing time for $1\times 10^7$ particle histories is reported to be around 2--30 s. The impressive speed gain compared to CPU-based calculations is due to both  algorithmic simplification and hardware acceleration. By far the most important approximation in refs.\cite{Kohno, Jia, Su, Osiecki} concerns proton-nucleus collisions. Su et al. \cite{Su} and Osiecki et al. \cite{Osiecki} have computed proton depth-dose curves in water on the GPU, without any nuclear interactions. Kohno et al. \cite{Kohno} indirectly consider nuclear processes by using measured proton depth-dose curves in water to determine energy losses during particle stepping. Jia et al. \cite{Jia} use a simplified model of non-elastic nuclear interactions (based on the work of Fippel and Soukup \cite{Fippel}), which can predict the energy spectrum of secondary protons from p--$^{16}$O collisions. The composition of all materials is assumed to be identical to water, and hence, only non-elastic collisions on $^{16}$O are considered. Their dose calculations in bone and low-density tissue demonstrate good agreement with Geant4, despite the lack of an in-depth model of nuclear interactions.

However, in certain situations a more accurate model of nuclear processes is critical, e.g. particle propagation in patients with sizeable metallic implants, %through high-Z materials such as metallic implants, 
detailed studies of secondary particle production, neutron dose estimates and dose-averaged linear energy transfer (LET$_{d}$) calculations.  A non-elastic proton-nucleus interaction can be simulated with the MC method as a two-step  process: a fast intranuclear (INC) cascade that leaves the nucleus in an excited state, followed by an evaporation stage \cite{Bertini}. Traditional CPU-based simulations are rather time-consuming: it takes $\sim$200s to compute $1\times10^6$ 200 MeV p--$^{16}$O non-elastic interactions\footnote{This is roughly the expected number of non-elastic events for $1\times 10^7$ proton histories.} with the Bertini INC model in Geant4.9.6p2 and the Li\`{e}ge INC model \cite{Liege} (INCL++ v5.1), respectively. In an effort to accelerate these calculations, we recently deployed a Bertini-style simulation of non-elastic interactions on the GPU \cite{INCPaper}, achieving a speed-up factor of $\sim$50 compared to the Geant4 Bertini model. 
%The energy and angular distribution of secondaries show good agreement with both Geant4 and experimental data. 

In this paper, we present a GPU-based class II condensed history proton transport MC, which uses our previously reported INC and evaporation kernels\footnote{In this paper, `kernel' refers to a piece of code that runs on the GPU but called from the host CPU.} \cite{INCPaper} to simulate non-elastic\footnote{The ICRU definitions for the terms `elastic', `non-elastic' and `inelastic' are adopted in this paper \cite{ICRU63}.} collisions with any nucleus, on an event-by-event basis. Elastic collisions of protons with nuclei other than hydrogen and oxygen are also modeled. Very good agreement is obtained with Geant4.9.6p2/TOPAS. On a NVIDIA GTX680 card, we report a net calculation time (including all CPU operations and CPU-GPU data transfers) of $\sim$20 s for a typical plan with $1\times10^7$ histories. Previous authors  \cite{Kohno, Jia, Su, Osiecki} report computational times that are of the same order. Compared with previous work, our simulation therefore contains a much more comprehensive treatment of nuclear processes, while achieving comparable net computational times. 
%The improved nuclear modeling also allows us to perform very fast LETd calculations on the GPU. This work opens exciting avenues for clinically-applicable MC-based IMPT optimization \cite{Ma}, as well as RBE-weighted treatment planning.
%We believe that QA of treatment plans with MC can be performed reliably, efficiently and cheaply, on a single computer with one graphics card. 

This paper is organized as follows. In \S\ref{section:simulationmodel}, we describe our physics model and all the major components of our simulation. This includes realistic beam spot modeling, energy loss, multiple scattering, energy straggling, INC, nuclear evaporation, elastic nuclear scattering and transport mechanics. In \S\ref{section:implementation}, the GPU implementation is described: our parallelization strategy (\S\ref{section:parallelization}), the software structure (\S\ref{section:softwarestructure}) and GPU memory management (\S\ref{section:gpumemory}). We give details of our validation methods and of our treatment plan re-calculation workflow in \S\ref{section:tpscalc} and \ref{section:validation} respectively. In \S\ref{section:results}, we report on the validation results, comparing our MC simulation results with Geant4.9.6p2/TOPAS: (1) predictions from our model of non-elastic nuclear interactions (\S\ref{section:nonelasticresults}), (2) pencil beam dose distributions in homogeneous phantoms (\S\ref{section:homogeneousphantoms}), and (3) MC re-calculations of complex head and neck treatment plans (\S\ref{section:treatmentplanresults}). We then discuss a number of current and future applications of our work in \S\ref{section:discussion}, before summarizing and concluding in \S\ref{section:conclusions}.

\section{Methods}
\subsection{Simulation model}\label{section:simulationmodel}
\subsubsection{Geometry}
We focus only on transport through voxelized geometries. Patient geometry is implemented from CT scans, using the formalism of Schneider et al. \cite{Schneider} to convert the Hounsfield Unit (HU) of each voxel to material density and composition. The HU to density conversion factors are the same as in ref.\cite{Schneider}. Following ref.\cite{Paganetti2008}, in the HU range between -1000 and 3060 we use 26 materials comprised of a combination of up to 12 elements, with everything above 3060 assumed to be  metallic implant (titanium by default).

\subsubsection{Stopping powers, multiple scattering, energy straggling}\label{section:dedxmultscat}
Total stopping power ($\mathrm{d}E/\mathrm{d}x$) tables for protons in water, air and titanium are obtained from Geant4.9.6p2. Proton range-energy tables are then generated via numerical integration in the continuous slowing down approximation (CSDA). Second derivatives are also pre-calculated so that the tables can be cubic-spline interpolated. The stopping power of a proton of energy $E$ in material $m$ with $-1000<\mathrm{HU}<3060$ is related to $S_w(E)$, the stopping power in water, by:
\begin{equation} S_m(E) = S_w(E) \frac{\rho_m}{\rho_w} f_m\end{equation}
where $\rho_{m}$ is the density of material, $\rho_w$ is the density of water, and $f_m$ is a scaling factor. $f_m$ is tabulated as a function of HU using Geant4.9.6p2. For air (HU = -1000) and titanium (HU $\ge$ 3060), this scaling factor is not applied, and pre-generated tables are used directly instead.  %We verified that it varies with energy within less than 0.1\%. 

Multiple scattering in each condensed history step is simulated using the Highland formula \cite{RPP}. A multiplicative factor of 1.07 is added to obtain comparable angular spread with Geant4.9.6p2, which employs the Urban multiple scattering model \cite{Urban}. Radiation lengths for all defined materials are pre-calculated with Geant4.9.6p2. We also implemented a full Moli\`{e}re multiple scattering simulation, which takes into account the single scatter tail at the cost of slowing down the simulation by a factor of $\sim$5. All results presented in this work were produced using the scaled Highland formula.

Energy straggling is simulated with a Vavilov-Gaussian model. Fast sampling from Vavilov distributions is achieved using Chibani's method \cite{Chibani}.  %The transition between the Vavilov and Gaussian regimes is determined by the parameter $\kappa$. 
The mean energy loss in each step is found by cubic-spline interpolating the range-energy table. For a material $m$ with $-1000 < \mathrm{HU} < 3060$, the CSDA proton range $R_m(E)$ is derived from the range in water, $R_w(E)$ by:
\begin{equation} R_m(E) = R_w(E) \frac{\rho_w}{\rho_m f_m}\end{equation}

In this work the kinetic energies of $\delta$-ray electrons are assumed to be deposited locally. Given the typical size of a voxel ($\sim$1 mm), this approximation holds very well for soft tissue and bone. 

\subsubsection{Transport mechanics}\label{section:transportmechanics}
The step size $\Delta x$ is initially picked as the minimum of three quantities: (1) the distance to the next elastic or non-elastic nuclear interaction, $\lambda$, which is selected according to the total nuclear cross-section  (2) the distance to the next voxel boundary, $d$, and (3) the range of the proton, $R(E)$. We use a `random hinge' algorithm \cite{PENELOPE} to take into account the small transverse displacement after each multiple scattering step. 

If $\Delta x = \lambda$, a nuclear collision occurs, and the proton is propagated to the collision point. A random number is thrown to determine if the collision is elastic or non-elastic. The proton history is terminated if one of three conditions are met: (1) $\Delta x = R(E)$, which means that the proton is stopping in the current voxel, (2) $E \le E_{cut}$, where $E_{cut} = 0.1$ MeV, and (3) the proton is outside the boundaries of the voxelized geometry.

\subsubsection{Non-elastic nuclear interactions}\label{section:nuclearmodel}
Non-elastic proton-nucleus cross-sections for all relevant nuclei in the energy range 0--250 MeV were tabulated using Geant4.9.6p2. Once a non-elastic collision is deemed to happen, a random number is thrown and the target is picked according to the relative sizes of the non-elastic cross-sections of each nuclear component. After picking a random point of entry into the target, the proton is propagated through nuclear matter. This is repeated until an INC involving at least one Pauli-allowed nucleon-nucleon collision occurs within the nucleus. We very briefly summarize the INC and evaporation simulations below (for further details see ref.\cite{INCPaper}).

The nucleus is assumed to be a two-component Fermi gas of neutrons and protons. The nucleon density follows a Woods-Saxon distribution that is approximated by 15 constant density shells. The incident proton and all collision products are tracked in the nuclear matter, until they exit the nucleus or become trapped. Pauli blocking is enforced by requiring the kinetic energies of all collision products to exceed their Fermi energies. Reflection and refraction at the potential shell boundaries are taken into account. In the therapeutic energy range ($E < 250$ MeV), only elastic collisions between nucleons need to be considered. Total and differential elastic nucleon-nucleon cross-sections are calculated using the parameterization of Cugnon et al. \cite{Cugnon}. The result of the INC simulation is a number of outgoing nucleons and a residual nucleus that subsequently de-excites by particle evaporation. 

To model the evaporation phase, we use the generalized evaporation model (GEM) \cite{GEM}. The angular distribution of evaporated particles is assumed to be isotropic in the rest frame of the nucleus. Other post-cascade processes such as photon evaporation, pre-equilibrium emission of particles, fission and Fermi break-up are currently not considered. Only protons produced in the INC and evaporation stages are stored for subsequent propagation. Neutrons are presently assumed to be dosimetrically irrelevant, and the kinetic energies of all charged particles heavier than protons are locally deposited.

\subsubsection{Coherent nuclear elastic interactions}\label{section:nuclearelastic}
We simulate coherent nuclear elastic collisions with all constituent nuclei in the patient geometry. Collision kinematics are fully relativistic. Total elastic cross-sections $\sigma_{el}$ are obtained from Geant4.9.6p2. Scattering angles $\theta$ in the center-of-mass frame are sampled according to the following parameterization of the elastic differential cross section by Ranft \cite{Ranft}: 
\begin{eqnarray}\nonumber 
\frac{\mathrm{d}\sigma_{el}}{\mathrm{d}\Omega} = A^{1.63}\mathrm{e}^{14.5\, t\,A^{0.66}} + 1.4A^{0.33}\mathrm{e}^{10t}\quad\quad A \le 62\\
\frac{\mathrm{d}\sigma_{el}}{\mathrm{d}\Omega} = A^{1.33}\mathrm{e}^{60\,t\,A^{0.33}} + 0.4A^{0.4}\mathrm{e}^{10t}\quad\quad\quad  A \ge 63
 \end{eqnarray}
where $A$ is the mass number of the target and the invariant momentum transfer $t$ in (GeV/c)$^2$ is given by 
$t = -2p^2(1-\cos\theta)$, $p$ being the momentum of the scattered proton in GeV/c.

\subsubsection{Simulation of proton beam spots from scanning nozzle}\label{section:phasespace}

\begin{figure}[!ht]\centering
\includegraphics[width=14.cm]{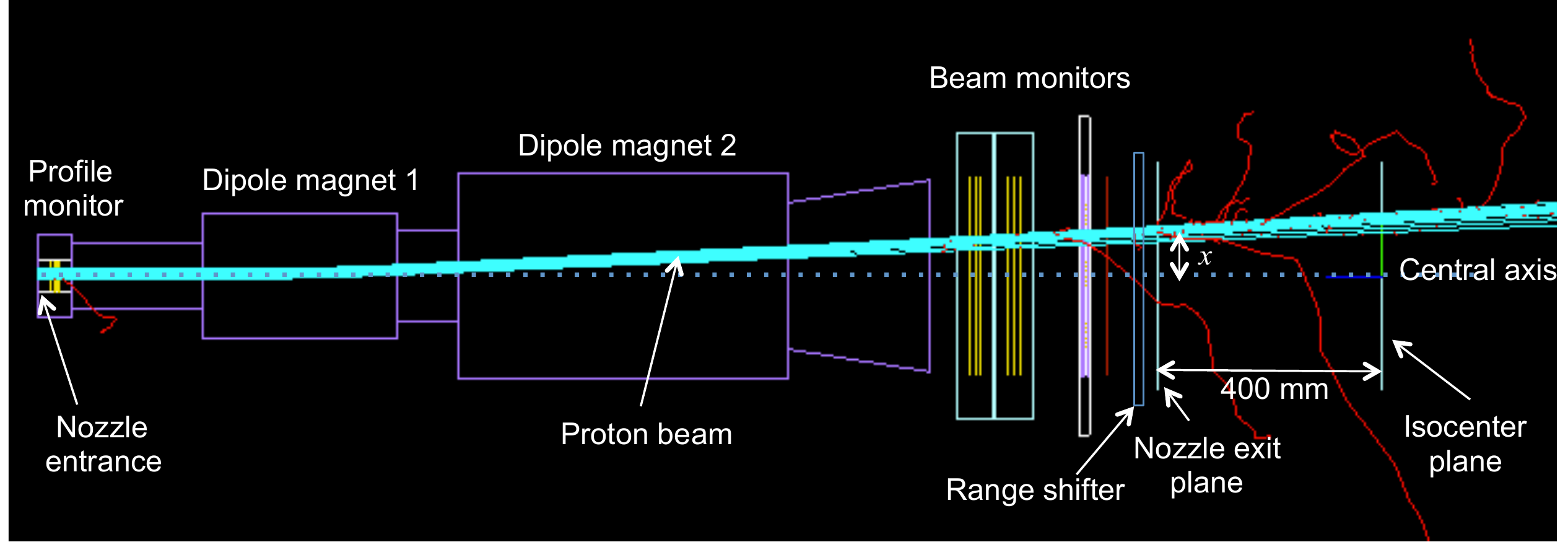}
\caption[]
 {The TOPAS nozzle model used to generate phase-space lists of primary protons that are read into our GPU-based MC.}
  \label{figure:nozzlepic}
 \end{figure}
 
A TOPAS simulation of the upcoming Mayo proton center spot scanning nozzles has been developed (fig.~\ref{figure:nozzlepic}). Details of the different treatment nozzle geometries and the proton beam phase space parameters at nozzle entrance were supplied by the vendor. Although commissioning data are not yet available to validate our model, preliminary spot size calculations are in good agreement with vendor specifications. The capability of Geant4 to accurately simulate spot scanning nozzles has previously been demonstrated in ref. \cite{MDAndersonNozzle}. 

We used the TOPAS simulation to generate phase-space lists of protons near nozzle exit at 40 cm from isocenter plane, covering all synchrotron beam energies and treatment nozzle configurations. Five kinematic variables are stored for each proton: the transverse positions $(x,y)$, transverse directional cosines $(u,v)$ and kinetic energy $E$. For any given beam energy, we verified that probability distributions of kinematic variables for all relevant off-axis beam-spots were essentially identical to those for central-axis spots. This significantly reduces the size of input phase-space data, since central-axis particle lists can be used to model off-axis beam spots. However, if the nozzle includes a range shifter, an energy reduction has to be applied to $E$ to account for the additional energy loss of protons in off-axis spots\footnote{For example, if the thickness of the range shifter is 4.34 cm, on average the beam at the extreme edge of the scanning region would `see' a thickness of 4.38 cm.}. This energy correction can be computed knowing the particle direction and the range shifter thickness.

The phase-space files are pre-loaded into our MC at initialization time, and protons are generated at the nozzle exit. In practice, the number of primary proton histories in a given calculation has to be scaled down to account for losses due to nuclear interactions in the nozzle; otherwise the dose deposited in the phantom or patient would be slightly overestimated. This scaling factor was calculated for each nozzle configuration and beam energy. The protons are then are propagated through air to the edge of the volume defined by the CT image in a single step, taking into account multiple scattering, energy straggling and nuclear interactions.

\subsection{GPU implementation}\label{section:implementation}
We now describe our implementation of the above model on a GPU using the CUDA C framework \cite{cuda}. 
The reader is assumed to be acquainted with GPU terminology; if not, he or she is referred to the abundant resources available on the NVIDIA developer website \cite{nvidia}.

%\subsubsection{The GPU environment}
%Describe:
%Threads, Warp, block, multi processor.
%Memory types: constant, shared, register, global, texture.
%Data transfers, memory reads, occupancy, divergence.

\subsubsection{Parallelization approach}\label{section:parallelization}
Many different strategies have been used in the past to parallelize photon, neutron and charged particle transport MCs on the GPU. Although no two simulations are identical, one can identify two broad paradigms: the `one step per thread', and `one particle history per thread' methods. 

In the first method (e.g. refs.\cite{Jahnke, rpi}), the simulation of particle histories is split into basic components (such as track stepping, boundary crossing and discrete physical interactions), which are accumulated and processed in parallel by kernels that contain little or no branching instructions. %Rare interactions may need to be accumulated before being simulated in batch, so as to achieve high GPU thread occupancy. 
As shown by Du et al. \cite{rpi} this strategy can result in a divergence-free simulation because all threads in a warp are made to execute the same set of instructions. 

The second method is the most commonly adopted approach. Here, a GPU thread processes a particle history until end conditions are satisfied. Since particle histories differ, thread divergence is inevitable, especially if more than one particle type are handled by a warp, or if a particle and its secondaries are tracked by the same thread. However, this method potentially results in a much reduced number of global memory transactions. For example, at the end of each step the particle position, direction and energy needs to be updated. These variables can be read from global to register memory, updated during track stepping and written back to global memory when end conditions are met. In contrast, in the `one step per thread' method, processing each track step would require at least one global memory read and write per kinematic variable, since one kernel launch occurs per step. 

Comparing the two methods for a neutron transport MC, Du et al. \cite{rpi} find the `one step per thread' method to be $\sim$10 times slower. In their implementation, the benefits of a divergence-free simulation is overwhelmed by severe global memory access latency. In this work we therefore follow the `one history per thread' approach. We limit the effects of thread divergence to some extent by  grouping non-elastic events and processing them in parallel. Our code structure is discussed next.

\subsubsection{Software structure}\label{section:softwarestructure}
We implement the following five kernels: (1) a phase space kernel to set the initial kinematic variables  of the protons in order to simulate realistic beam spots (\S\ref{section:phasespace}), (2) a transport kernel to evolve the proton history step by step through the voxelized geometry, taking into account energy loss fluctuations and scattering (\S\ref{section:dedxmultscat}), until it undergoes a non-elastic collision or meets one of the end conditions outlined in \S\ref{section:transportmechanics}, (3) an INC simulation kernel, (4) a nuclear evaporation kernel, and (5) a dose statistics kernel to evaluate the total and root mean square (RMS) dose in each voxel. Kernels (3) and (4) are described in further detail in ref.\cite{INCPaper}.
We also use CURAND kernels \cite{curand} for on-the-fly random number generation, and CUDA Thrust \cite{Thrust} libraries for various bookkeeping operations.

\begin{figure}[]\centering
\includegraphics[width=9cm]{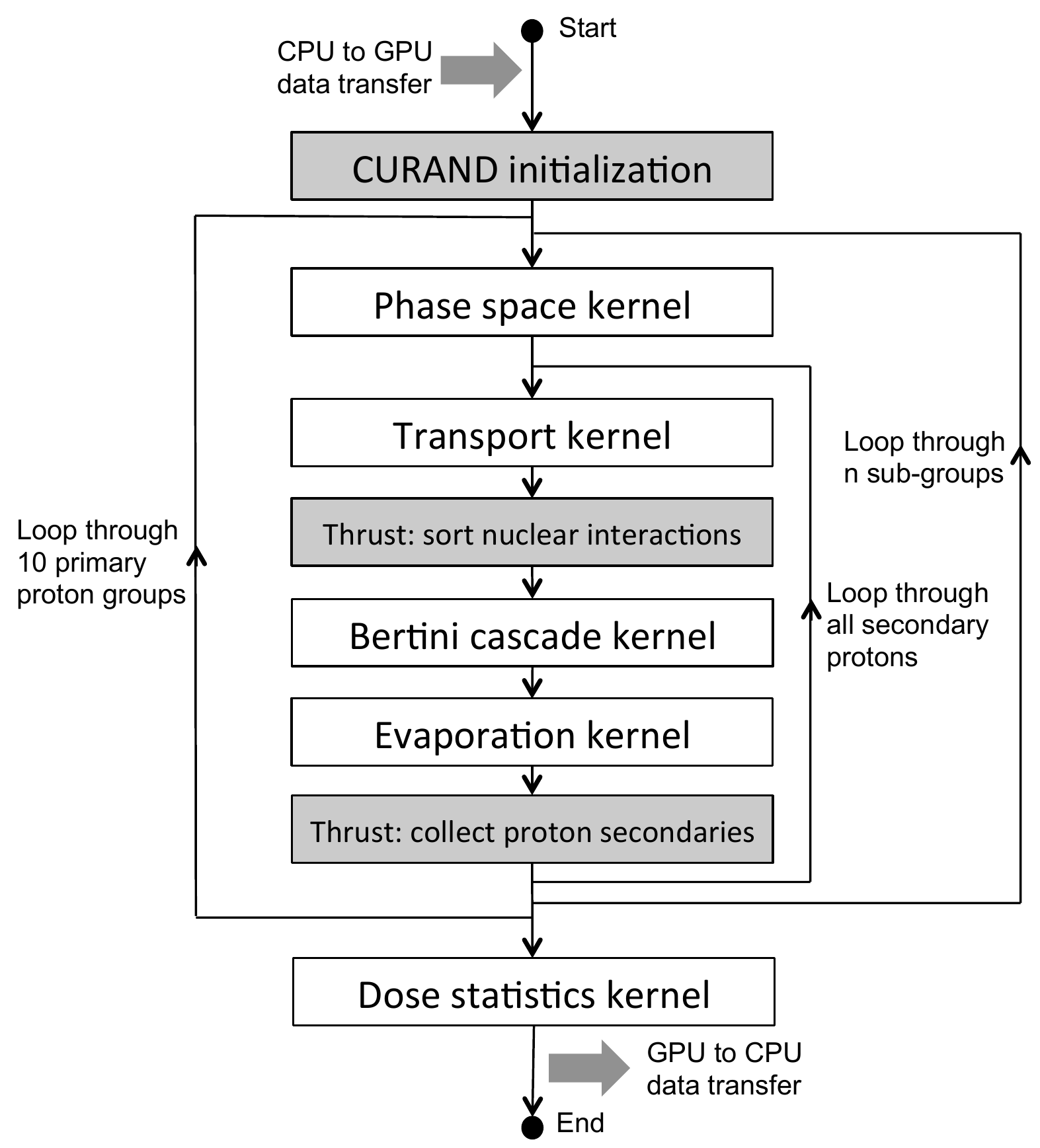}
\caption[]
 {The sequence of kernel calls in our simulation.}
  \label{figure:spockflowchart}
 \end{figure}

For a dose calculation involving $N$ proton histories, we process $10$ groups of $N/10$ protons to evaluate statistical errors on the dose deposited in each voxel. Due to GPU memory constraints, a group is further split into sub-groups of $n$ protons, where $n$ is an integer multiple of the number of GPU threads in a block. 

The sequence of kernel calls is shown in fig.~\ref{figure:spockflowchart}. Before looping through the primary proton groups, the voxelized geometry, CURAND states, cross-sections, stopping power and range tables, physics constants, dose array and other simulation inputs are initialized. At the start of a proton sub-group loop, the phase space kernel is launched to sample the kinematic variables of  all primary protons in that sub-group. Then the transport kernel is launched to simulate all protons until end conditions are met. If a nuclear interaction occurs, it is stored for subsequent batch processing. Thrust utility kernels are then called to collect and sort non-elastic interactions. These are simulated in parallel by the INC kernel, resulting in a list of residual nuclei. This list is passed to the evaporation kernel to simulate the de-excitation mechanism. The INC and evaporation kernels are based on the model in \S\ref{section:nuclearmodel}. Secondary protons produced in the transport, INC and evaporation kernels are collected using Thrust functions, and the three kernels are again invoked in the same order. This is repeated until all secondary protons are simulated. The program then moves on to the next primary proton group. After all groups have been processed, the dose statistics kernel calculates the total and RMS dose in each voxel from the dose tallies of each primary proton group.

In our MC, CPU operations are limited to simulation initialization, kernel call control, and file output. The thread workload for each GPU kernel is as follows: the phase space kernel initializes the position $(x, y, z)$, directional cosines $(u, v, w)$ and kinetic energy $E$ of one proton in each GPU thread. The transport kernel handles one proton history per thread, and the Bertini cascade and evaporation kernels simulate one INC and one nuclear de-excitation per thread, respectively. As for the dose statistics kernel, it processes one voxel per thread.

\subsubsection{GPU memory allocation and management}\label{section:gpumemory}
Read-only data such as phase space kinematic variables, interaction cross-sections and stopping power and range tables are stored in as 1-D textures. Constant memory is used for storing physics and mathematical constants. Shared memory is barely used in this work.  CT geometry information and the 3-D dose tallies of each primary proton group are kept in GPU global memory. All dose tallies need to be updated `atomically' to prevent race conditions when multiple threads are updating the dose deposited to the same voxel. For any proton group, a single dose tally was found to be sufficient on Kepler cards\footnote{On Fermi cards, Jia et al.  \cite{Jia} report the calculation speed to be considerably extended by atomic operations. They use multiple dose tallies to mitigate this effect.}. Global memory is also used to store stacks of particles and excited nuclei, using structures of arrays (SoAs) to group together information relevant to each entity (e.g. a SoA of particles contains pointers to position, direction and kinetic energy arrays on device memory). Using SoAs guarantees coalesced global memory transactions. To minimize global memory traffic, SoAs are read and updated once per kernel call. For example, in a transport kernel call, information contained in the particle SoA is read into register memory, modified many times during particle tracking and written back to the SoA at the end of the call. 

As seen in fig.~\ref{figure:spockflowchart}, virtually all host-GPU data transfers occur either during the initialization stage or at simulation end, when the computed dose and dose RMS maps are transferred to the host CPU for file output. %For example, CT geometry data, cross-section and phase space cumulative probability tables are created and transferred to the GPU before all kernel calls. 
Particle SoAs are created and cleared at the beginning and end of a sub-group loop, respectively. Data contained in the SoAs do not need to be transferred to the host between kernel calls. In other words, we tried to eliminate CPU-GPU data transfers as much as possible.

\subsection{Treatment plan calculations}\label{section:tpscalc}

\begin{figure}[!htp]\centering
\includegraphics[width=12cm]{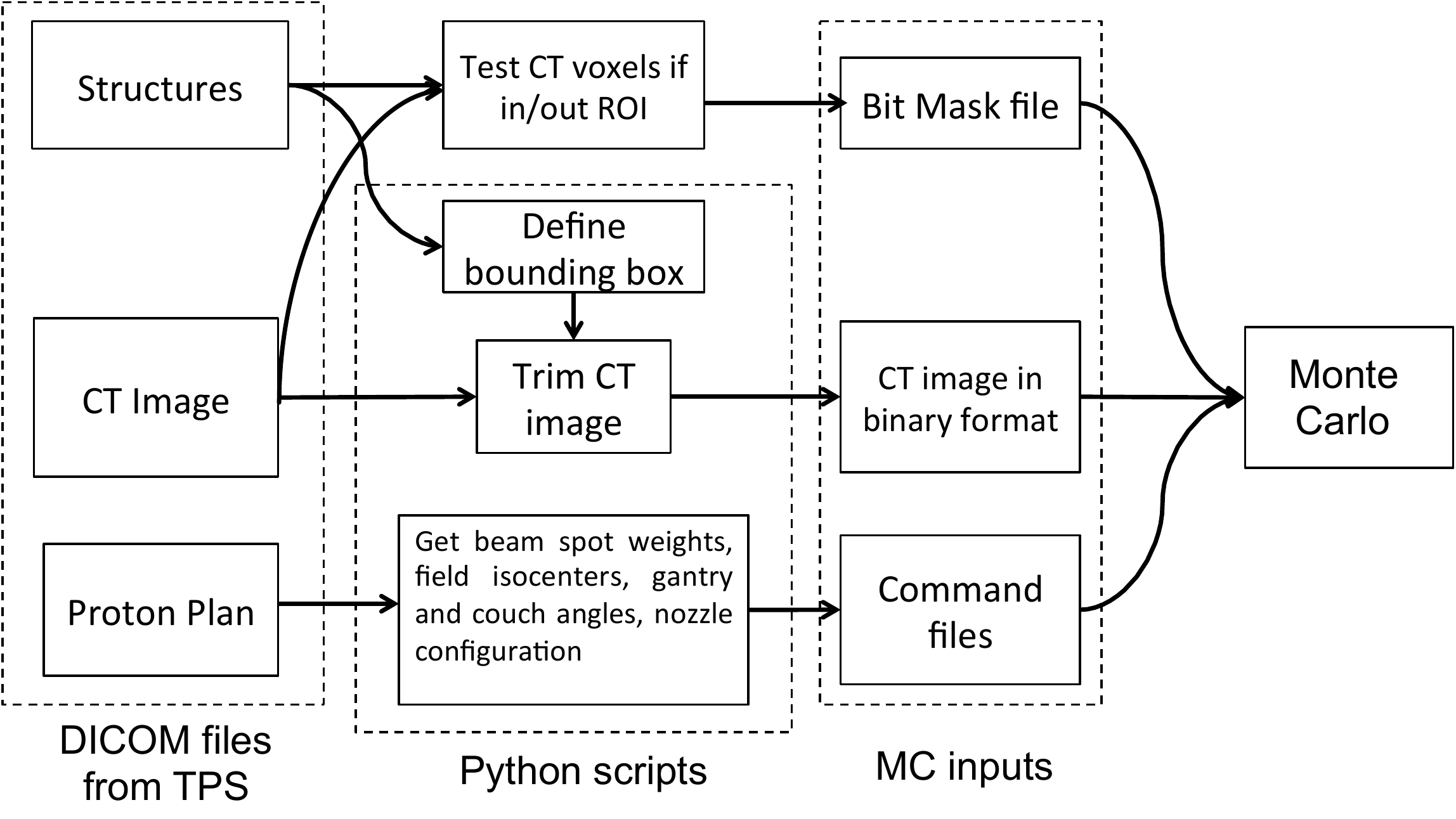}
\caption[]
 {Flow diagram showing pre-processing steps for simulating plans from TPS.}
  \label{figure:dosecalcflowchart}
 \end{figure}

The workflow that we adopted for recalculating plans from a commercially available TPS is shown in fig.~\ref{figure:dosecalcflowchart}. DICOM files (plan, structure and image) outputted by the TPS are processed using a set of Python scripts, to produce inputs for our MC. This step involves trimming the CT image to reduce computational time: using information contained in the structures file, a smaller rectangular volume containing the patient geometry is defined, so that air-filled voxels in the beam path are removed as much as possible. Beam spot weights, isocenter positions, gantry and couch rotation angles, and details of the nozzle configuration are also read and outputted in text format. A `bit mask' file specifying whether or not each voxel in the reduced volume is within a user-specified list of structures is created. This information allows the user to override the HU values of all voxels inside or outside a given structure, and to restrict dose file output to certain structures only. This considerably reduces GPU-CPU data transfer and file output times at simulation end.

\subsection{Validation}\label{section:validation}
To validate our nuclear model, we simulated non-elastic proton collisions on a large number of  therapeutically relevant nuclei, with incident energies between 70 and 230 MeV. We compared our predictions with Geant4.9.6p2 (using the Bertini and Binary cascade models) and published measurements when available. 

Using our MC, we simulated infinitesimally narrow pencil beams of protons with energies between 70 and 230 MeV, in various homogeneous voxelized phantoms including water, soft tissue, dense bone and pure calcium and titanium. The voxel size was 1 mm along all directions. The simulations were also carried out with Geant4.9.6p2, using the QGSP\_BIC\_EMY \cite{qgspbicemy} physics list. All particles were terminated once they exit the phantom, so as to eliminate any in-scattering contributions from outside the region of interest. These pencil beam simulations allowed us to verify and refine our implementation of the physics models for proton transport. We also simulated square fields of varying sizes in water at several beam energies, for different treatment nozzle configurations. These square field simulations were repeated with  TOPAS, using the nozzle model described in \S\ref{section:phasespace}. 

To verify our treatment plan calculations, we simulated three complex head and neck cases involving different patients using both TOPAS and our GPU-based MC. The TOPAS simulation again included the nozzle model described in \S\ref{section:phasespace}. Both simulations adopted the native CT voxel size ($1.25\times1.25\times2.5$ mm$^3$) for particle transport and dose scoring. To quantify differences between the two MC we use the 3-D gamma index \cite{Low}, which is also computed on the GPU. However, the 3-D gamma index might not be sensitive to systematic biases that are smaller than the pass criterion. To investigate these, we examined distributions of the percentage dose difference, defined for a given voxel as:
\begin{equation}\Delta_{G4-GPU} = 100\cdot \frac{D_{GPU}-D_{G4}}{D_{G4}} \end{equation}
where the dose calculated by TOPAS and our MC in that voxel are $D_{G4}$ and $D_{GPU}$, respectively. We also define the quantities $\Delta_{G4-G4}$ and $\Delta_{GPU-GPU}$. The latter is defined as:
\begin{equation}\label{equation:dgpugpu}\Delta_{GPU-GPU} = 100\cdot \frac{D_{GPU}-D'_{GPU}}{D'_{GPU}} \sim \sqrt{2}\sigma_{stat}^{GPU} \end{equation}
where $D'_{GPU}$ is the dose calculated by our MC for the same number of protons, but using a different starting random number seed, and $\sigma_{stat}^{GPU}$ is the average percentage statistical uncertainty on the dose in the voxel. $\sigma_{stat}^{GPU}$ is estimated by dividing the total number of primary protons in 10 groups, as explained in \S\ref{section:softwarestructure}. $\Delta_{G4-G4}$, which measures the statistical error in the TOPAS simulation, is defined in a similar way.

Ideally, 1-D distributions of $\Delta_{G4-GPU}$, $\Delta_{G4-G4}$ and $\Delta_{GPU-GPU}$ in the targets  should be Gaussians peaked at zero, with identical widths. Biases and other model discrepancies result in peak shifts and RMS differences.

\section{Results}\label{section:results}
\subsection{Non-elastic interactions}\label{section:nonelasticresults}
\begin{figure}[!ht]\centering
\includegraphics[width=7.75cm]{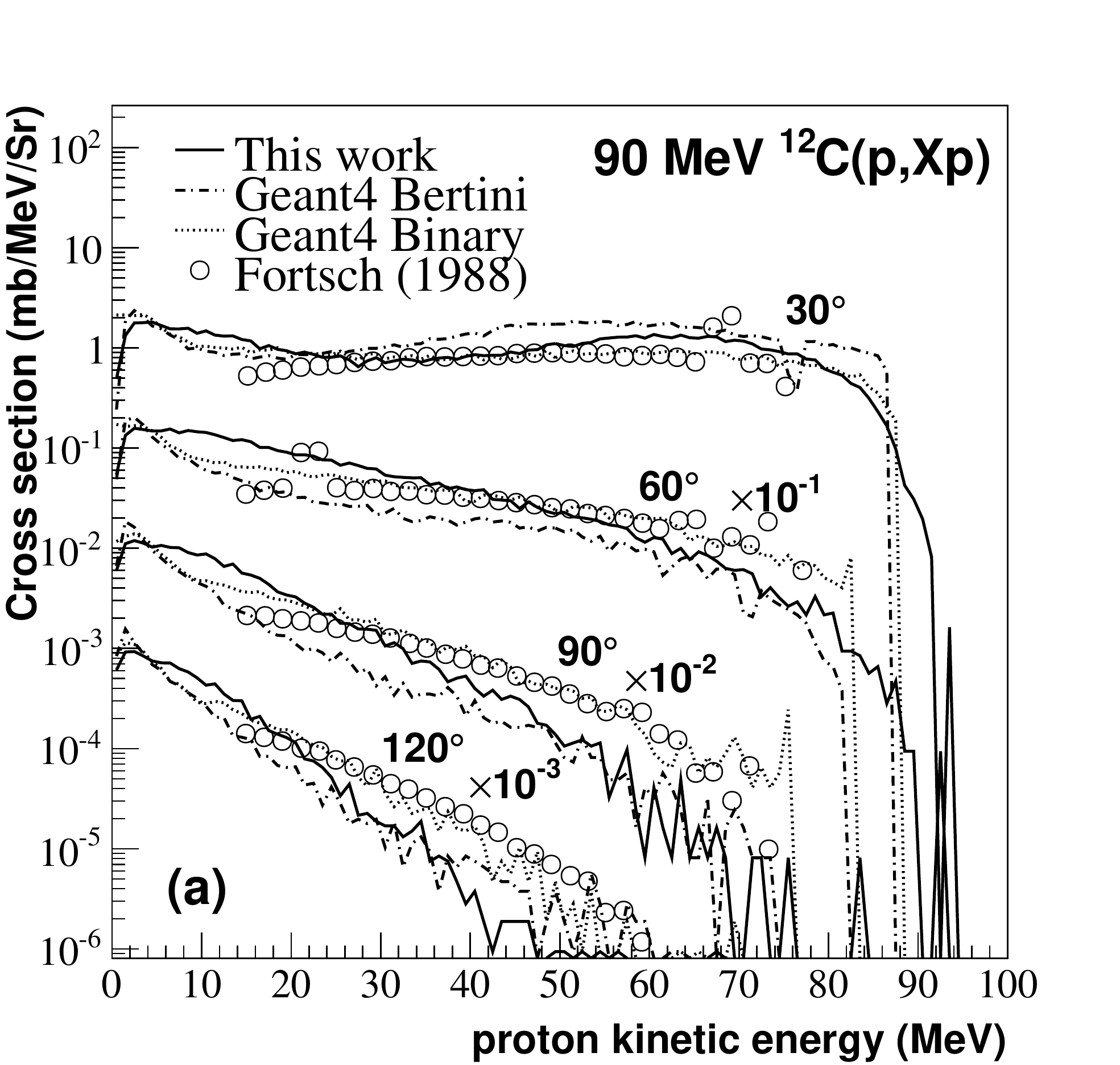}
\includegraphics[width=7.75cm]{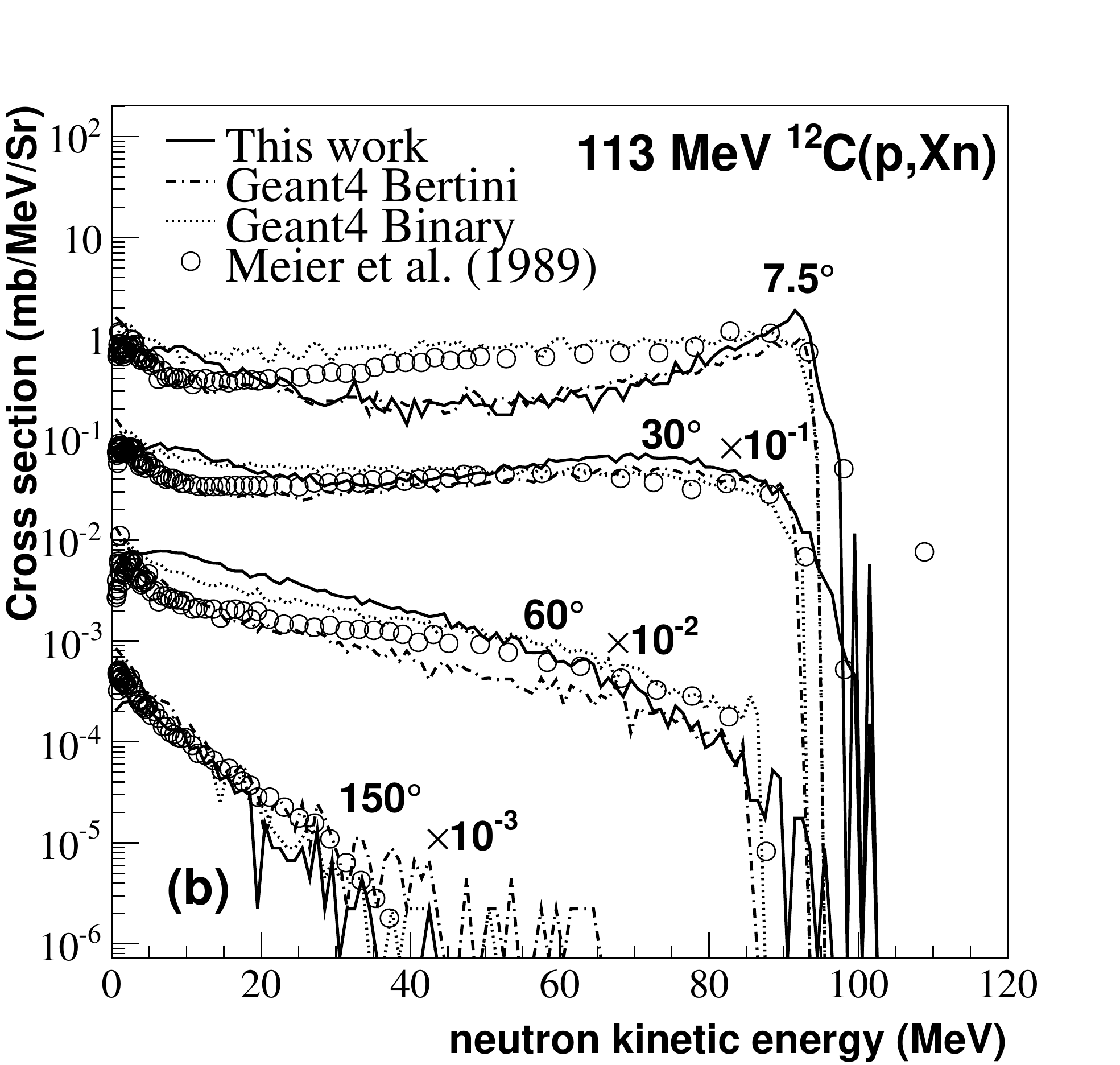}
\includegraphics[width=7.75cm]{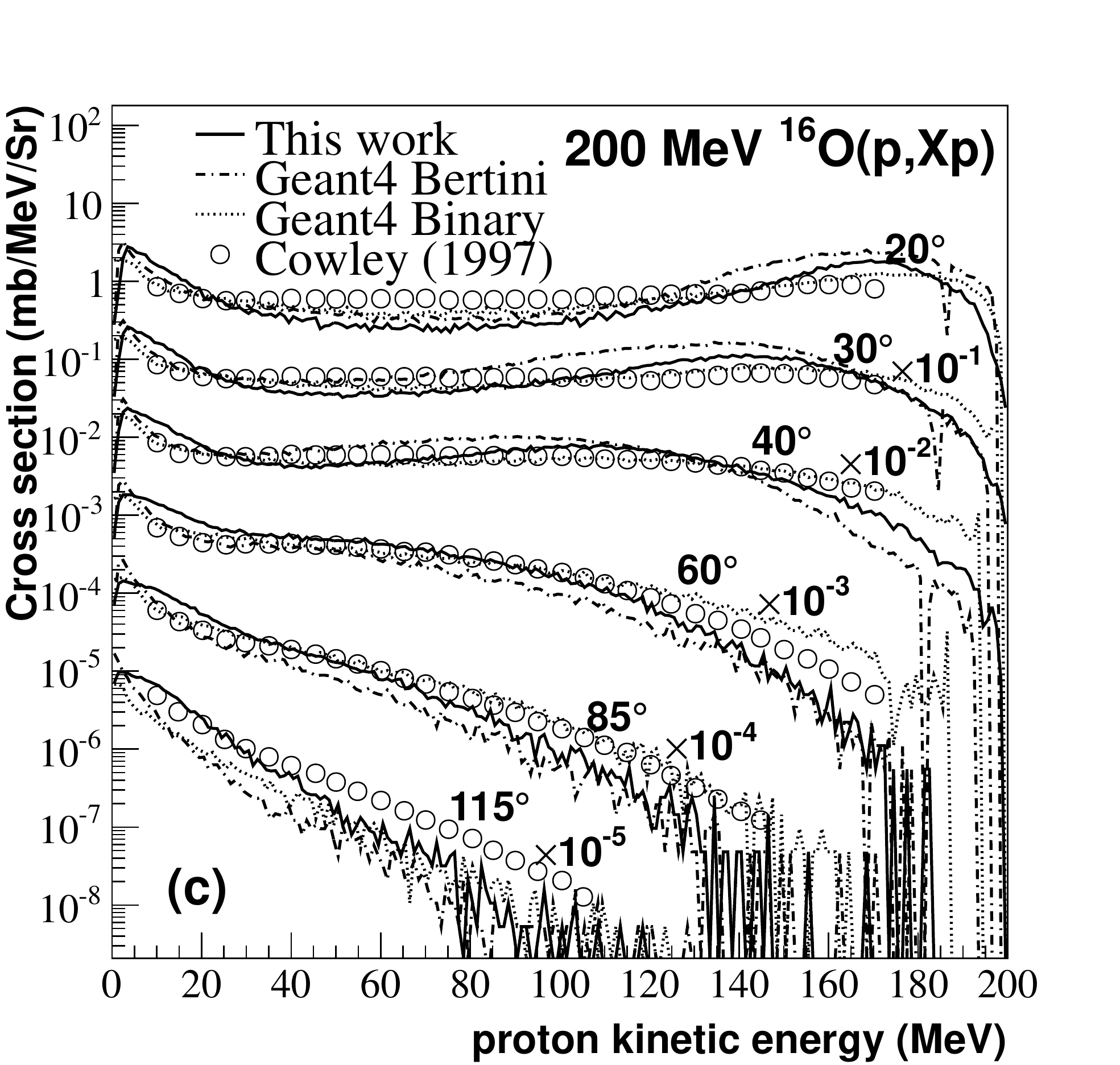}
\includegraphics[width=7.75cm]{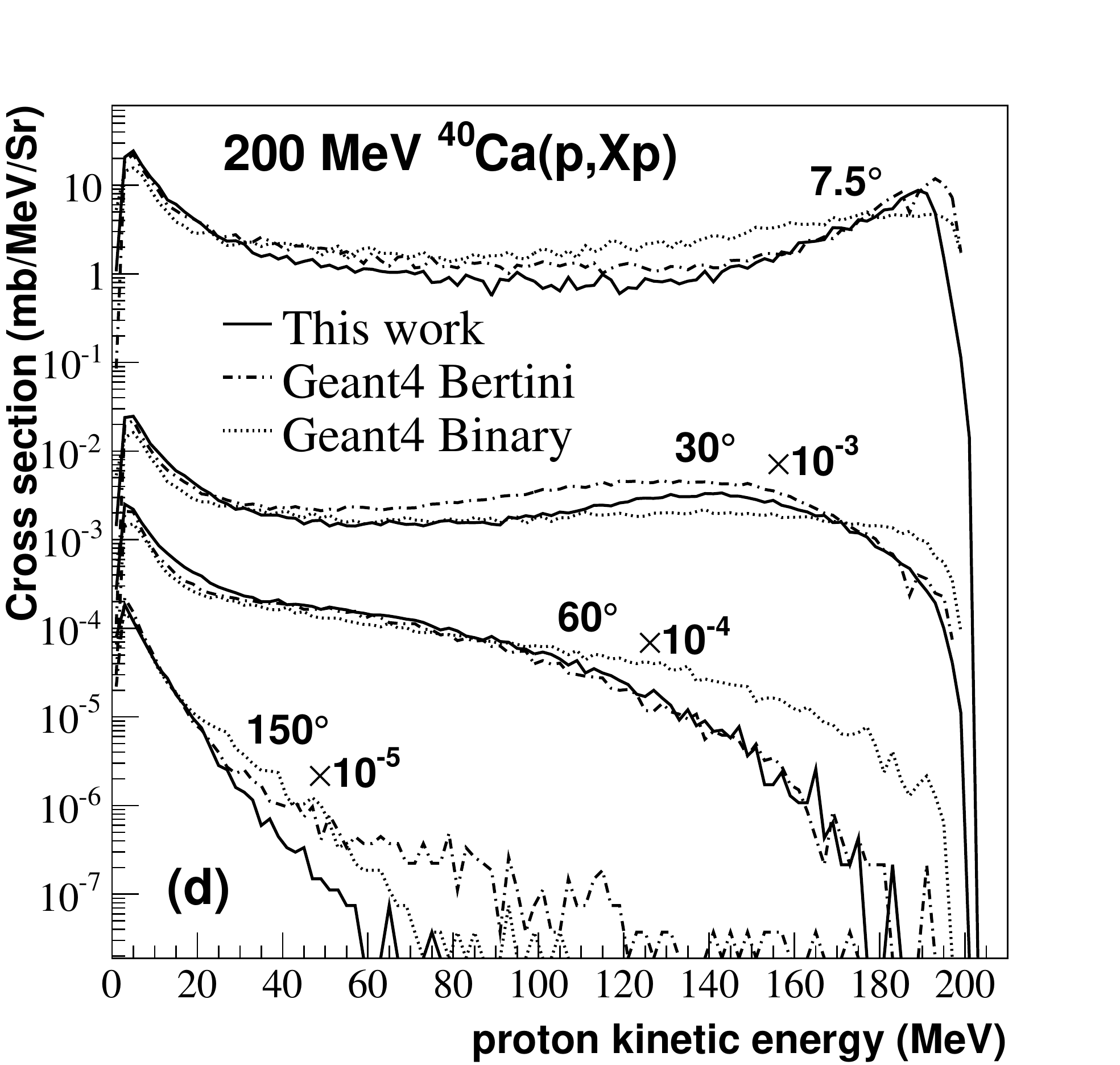}
\caption[]
 {Calculated double-differential cross-sections for secondary neutron and proton production in (a) 90 MeV p--$^{12}$C, (b) 113 MeV p--$^{12}$C, (c) 200 MeV p--$^{16}$O and (d) 200 MeV p--$^{40}$Ca collisions, compared with Geant4 Bertini and Binary cascade model predictions and measurements from refs.\cite{Fortsch, Meier, Cowley}.}
  \label{figure:incCalcs}
 \end{figure}

Fig.~\ref{figure:incCalcs} shows predicted differential cross-sections for proton and neutron production in: (a) 90 MeV p--$^{12}$C, (b) 113 MeV p--$^{12}$C, (c) 200 MeV p--$^{16}$O and (d) 200 MeV p--$^{40}$Ca collisions at various angles. The values are normalized to total cross-sections from Cugnon et al. \cite{Cugnon} and have been scaled down for display clarity (except for the topmost curves). In general, reasonable agreement is obtained with both Geant4.9.6p2 models and published data. The Bertini and Binary models are routinely included in the physics lists used in Geant4-based dose calculations. Further comparisons of our INC and evaporation kernel predictions with Geant4 are shown in ref.\cite{INCPaper}. 

\begin{table}[hpt]
\begin{center}
\footnotesize\rm
\begin{tabular}{@{}ccccc}
\hline Platform & Model & p--$^{16}$O (s) &p--$^{40}$Ca (s)\\\hline\hline
(1) CPU& G4 Binary &2382.07&4620.92\\
(2) CPU& G4 Bertini &492.42&534.13\\
(3) CPU &this work&172.74&326.11\\
(4) GPU &this work&8.39&15.56\\
(5) GPU &this work&6.61&9.77\\\hline
\end{tabular}
\caption{\label{table:INCruntimeresults} Time taken in seconds to compute $2.6\times 10^6$ proton--$^{16}$O and proton--$^{40}$Ca interactions at 200 MeV incident proton kinetic energy (K.E).}
\end{center}
\end{table}

Table \ref{table:INCruntimeresults} shows the computational times for $2.6\times10^6$ proton--$^{16}$O and proton--$^{40}$Ca interactions at 200 MeV for the following: (1) Geant4 Binary cascade model on the CPU, (2) Geant4 Bertini cascade model on the CPU, (3) The CPU version of our MC, (4) Our GPU-based MC, and (5) Our GPU-based MC, compiled with the \texttt{-use\_fast\_math} flag\footnote{Compiling with the \texttt{-use\_fast\_math} flag enables the use of faster, but less accurate functions for standard mathematical operations on NVIDIA GPUs.}. We used a i7-3820 3.6 GHz processor for the CPU tests, and the GPU calculations were performed on a single NVIDIA GTX680 card. It is seen that our MC can be $\sim$50 times faster than the Geant4 Bertini model. It was verified that identical results were obtained in cases (3), (4) and (5).

\subsection{Calculations in homogeneous phantoms}\label{section:homogeneousphantoms}

\begin{figure}[]\centering
\includegraphics[width=12cm]{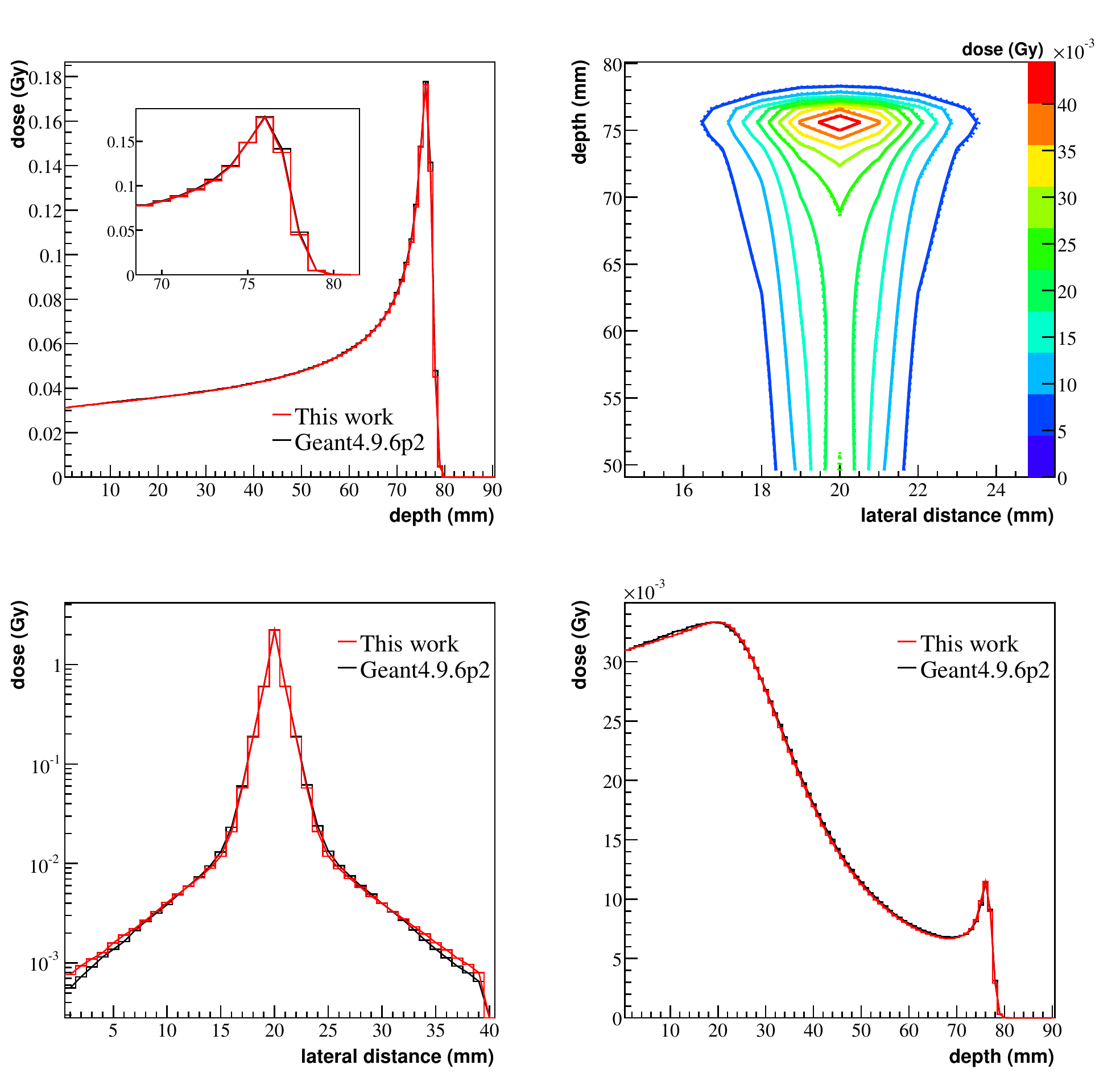}
\caption[] 
{Clockwise from top left: our calculated depth-dose, 2D iso-dose contours, transverse dose and central axis depth-dose distributions for a pencil beam of 100 MeV protons in water, compared with Geant4.9.6p2. In the top left figure, the inset is a zoom around the Bragg peak. In the top right figure, the iso-dose contours from our simulation are displayed in solid lines, while the Geant4.9.6p2 contour lines are dotted (not visible because of the overlap).}
  \label{figure:pencilbeam100MeVwater}
 \end{figure}
 
 \begin{figure}[]\centering
\includegraphics[width=12cm]{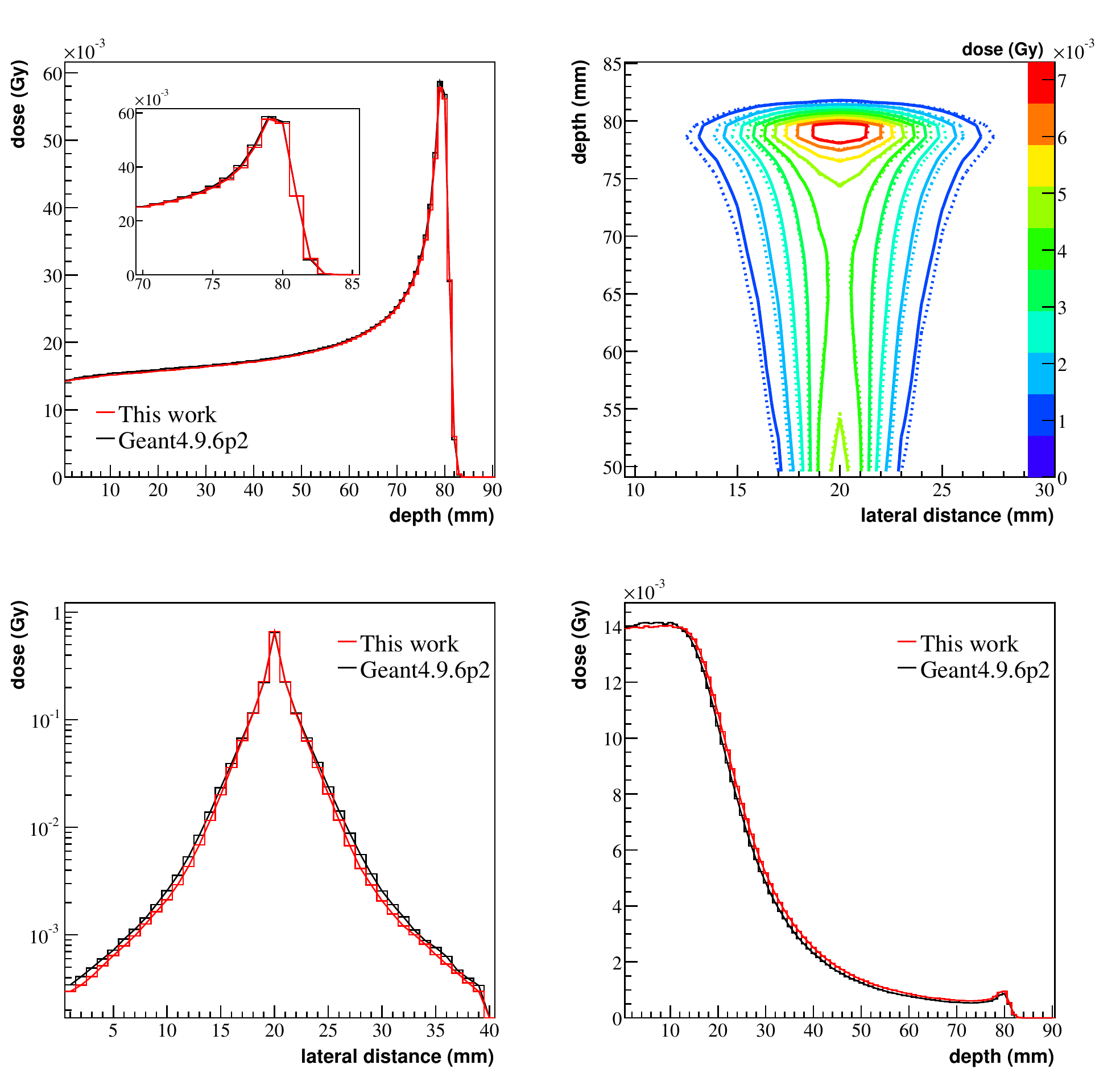}
\caption[]
{Same as fig.~\ref{figure:pencilbeam100MeVwater}, but for 200 MeV protons in pure titanium.}
  \label{figure:pencilbeam200MeVTi}
 \end{figure}

% \begin{figure}[]\centering
%\includegraphics[width=12cm]{paperplot230MeVHU3000.pdf}
%\caption[]
%{Same as in fig.~\ref{figure:pencilbeam230MeVHU3000}, but for 230 MeV protons in dense bone.}
%  \label{figure:pencilbeam230MeVHU3000}
 %\end{figure}

To illustrate some of the pencil beam simulation results, figs.~\ref{figure:pencilbeam100MeVwater} and \ref{figure:pencilbeam200MeVTi} show, clockwise from the top left: the depth dose, 2-D isodose contours, transverse dose and central-axis depth-dose profiles for $2.6\times10^6$ 100 and 200 MeV protons in water and titanium, respectively.  In these two figures, the average statistical error on the dose in voxels containing at least 10\% of the maximum dose was below 0.2\%. Very good agreement was obtained with Geant4.9.6p2: using the Geant4 dose maps as the reference, the 3-D gamma pass rate at 2\%-2 mm was 100\% for all beam energies and phantoms investigated. In particular, our MC correctly models the transverse dose tails, down to the regime where nuclear elastic scattering dominates. Fig.~\ref{figure:pencilbeam200MeVTi} explicitly demonstrates the capability of our MC to correctly model proton transport in a typical metallic implant. We stress that the level of agreement illustrated fig.~\ref{figure:pencilbeam200MeVTi} would have been difficult to achieve with a nuclear model that only considers p--$^{16}$O interactions. 

The square field simulations also demonstrated very good agreement between Geant4 and our MC. These calculations were performed to test our ability to compute large field sizes, and to correctly position and model beam spots. For brevity, we won't show these results, since the treatment plan recalculation results below imply that these requirements were met.

%\subsection{Square field calculations}

\subsection{Treatment plan calculations}\label{section:treatmentplanresults}

 \begin{figure}[!ht]\centering
\includegraphics[width=12cm]{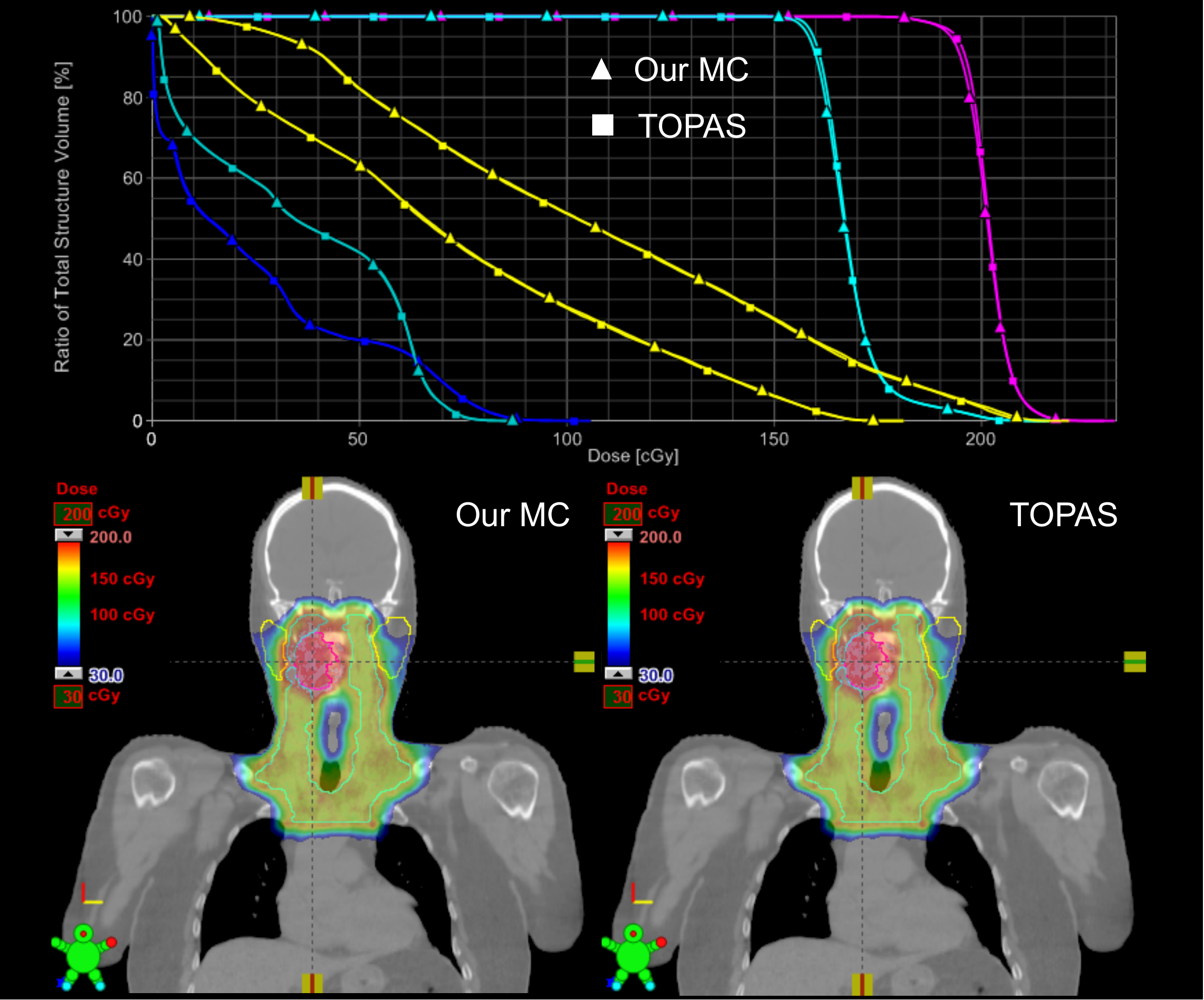}
\caption[]
{Re-calculation of a two-target, three-field head and neck plan from a commercially-available TPS using our MC and TOPAS. Top: comparison of DVHs. The curves are, from left to right: cord, brain stem, left parotid, right parotid, low-dose CTV and high-dose CTV. Bottom left and right: dose color wash in a representative coronal plane from our MC and TOPAS, respectively.}
  \label{figure:barfigpaper}
 \end{figure}

\begin{figure}[!ht]\centering
\includegraphics[width=14cm]{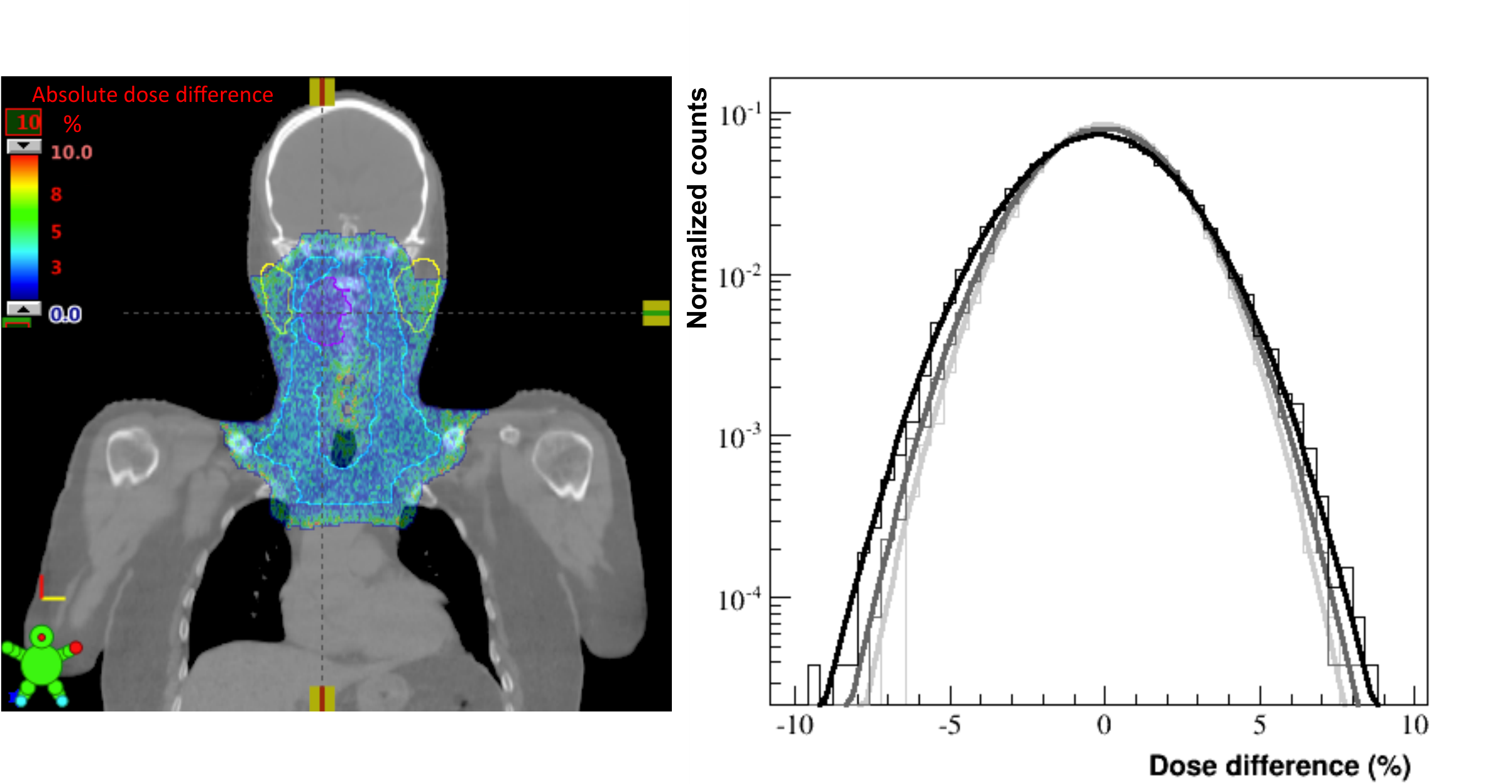}
\caption[]
{ Left: absolute percentage dose difference between TOPAS and our MC, $|\Delta_{G4-GPU}|$, for the  coronal plane shown in fig.~\ref{figure:barfigpaper}. Right: Gaussian fits to distributions of $\Delta_{G4-GPU}$ (black), $\Delta_{G4-G4}$ (dark grey) and $\Delta_{GPU-GPU}$ (light grey) for all voxels within the high-dose CTV. }
  \label{figure:bar_dose_diff}
 \end{figure}

Fig.~\ref{figure:barfigpaper} shows MC re-calculations of one of the three head and neck plans from a commercially available TPS. This plan consisted of two bilateral and one posterior fields. The dose in one representative coronal plane is shown in the bottom plots for our MC (left) and TOPAS (right). This plan contained one low-dose and one high-dose target that are delineated in cyan and magenta, respectively. The top plot in fig.~\ref{figure:barfigpaper} shows predicted DVHs for the two targets and four critical structures, for our MC (triangle markers) and TOPAS (square markers). Very good agreement is obtained between the two simulations. Taking the TOPAS dose map as the reference, the 3-D gamma pass rate at 2\%-2 mm in this case was 98.2\%. 
A color wash plot showing the distribution of $|\Delta_{G4-GPU}|$ in the same coronal plane is shown  in fig.~\ref{figure:bar_dose_diff} (left). No apparent pattern in $|\Delta_{G4-GPU}|$ was found; as expected higher values were obtained in voxels with low dose statistics. 

Distributions of $\Delta_{G4-GPU}$, $\Delta_{G4-G4}$ and $\Delta_{GPU-GPU}$ in the high-dose target are Gaussian, as shown in the right plot in fig.~\ref{figure:bar_dose_diff}. 
We observe that the distribution of $\Delta_{G4-GPU}$ is slightly off-center, i.e. $D_{GPU}$ is on average 0.2\% lower than $D_{G4}$. This small dose deficit is attributed primarily to neutrons and de-excitation gammas originating from nuclear interactions in the nozzle and in the patient, and which are not currently propagated in our MC simulation. Gaussian fits to the three curves yield RMS values of 2.2, 2.0 and 1.9, respectively. The value of 1.9 is consistent with the estimated value of $\sigma_{stat}^{GPU} = 1.3$\% (see Eq.~\ref{equation:dgpugpu}). The percentage statistical error in the TOPAS simulation is slightly higher ($\sigma_{stat}^{G4} \sim \sqrt{2}$\%). This is possibly caused by the difference in energy straggling implementations between TOPAS and our MC. Note that in our simulation the initial number of primary proton histories has been scaled down relative to TOPAS to account for nuclear interactions in the nozzle (see \S\ref{section:phasespace}). 

$\sigma_{total}^2$, the variance of the distribution of $\Delta_{G4-GPU}$,  can be written as:
\begin{equation} \sigma_{total}^2 = (\sigma_{stat}^{GPU})^2 +  (\sigma_{stat}^{G4})^2 + \sigma_{model}^2\end{equation}
where $\sigma_{model}^2$ is the additional variance due to model differences. From the above numbers, $\sigma_{model}\sim 1$\%. We hypothesize that the main contribution to $\sigma_{model}$ is due to the handling of $\delta$-rays: TOPAS propagates these particles, while in our simulation the kinetic energy of electrons is locally deposited.

For the two other head and neck plan re-calculations, the 3-D gamma pass rates at 2\%-2mm were 97.8\% and 98.9\%. Very good agreement was again obtained between the DVHs generated from our MC and TOPAS, and dose difference distributions followed the same trend. Hence, for brevity we will not discuss these two plans in further detail.

The computational times for the three head and neck plans on two types of NVIDIA cards are shown in table \ref{table:runtimeresults}. We verified that simulations performed with and without the \texttt{-use\_fast\_math} flag were equivalent. It is seen that our MC on a single K20Xm card can be up to $\sim$200 times faster than TOPAS on a 100-node cluster. Evidently, TOPAS performs a more detailed calculation, but as shown above, dosimetrically the results from the two simulations are very close.

\begin{table}[hpt]
\begin{center}
\footnotesize\rm
\begin{tabular}{@{}ccccc}
\hline Platform & HN Plan 1 & HN Plan 2 & HN Plan 3\\\hline\hline
Our MC (K20Xm \texttt {-use\_fast\_math}) & 119 s & 128 s & 129 s \\
Our MC (GTX680  \texttt {-use\_fast\_math}) & 155 s & 165 s & 176 s\\\hline
%This work (M2090  \texttt {-use\_fast\_math})& 25 s&& 163 s\\
%This work (C2050  \texttt {-use\_fast\_math})& 26 s&& 193 s\\\hline
Our MC (K20Xm) & 225 s & 239 s & 270 s\\
Our MC (GTX680) & 256 s &270 s &  288 s \\\hline
%This work (M2090) & 39 s&& 327 s\\
%This work (C2050) & 46 s& &412 s\\
%This work (laptop - GT650M) &111 s&& 1149 s\\\hline
 TOPAS on CPU cluster && $\sim$650 CPU-hours &\\\hline
\end{tabular}
\caption{\label{table:runtimeresults} Time taken by a single NVIDIA K20Xm and GTX680 card to compute three different head and neck plans, each with $6\times 10^7$ proton histories. }
\end{center}
\end{table}

\section{Discussion}\label{section:discussion}

\subsection{Current applications of our GPU-based MC}

At our clinic, the GPU-based proton transport MC is being used in two applications that can potentially play an important role in mitigating the effects of proton range uncertainties:

\begin{enumerate}
\item Our MC was deployed on a dual K20Xm GPU server to routinely QA treatment plans from a commercially-available TPS. Due to the very fast computational speed, near real-time feedback on the accuracy of most plans is achievable. This would not have been possible with a conventional, CPU-based MC. A friendly user interface for this verification system is being developed to make the software easily accessible to dosimetrists. Another Python script, not shown in fig.~\ref{figure:dosecalcflowchart}, allows the MC-calculated dose to be imported into the TPS for convenient  evaluation and display. 

\item In addition, we have developed a GPU-based intensity modulated proton therapy (IMPT) optimization code that adopts our MC as the dose computation engine \cite{Ma}. For a complex three-field head and neck plan, the full calculation, including the initial dose map for all relevant beam spots, is estimated to take half an hour on a cluster with 24 GPUs. Therefore, using GPU acceleration, MC-based IMPT planning can be clinically applicable on a routine basis. We are currently upgrading this code to a robust MC-based IMPT optimization system. 
\end{enumerate}

\subsection{Future applications}
We are planning the following two improvements to our physics model: simulation of the production and transport of $\delta$-rays, and propagation of secondary neutrons from non-elastic nuclear interactions of protons in the patient. The ability to evolve neutron histories will allow us to make very fast and accurate neutron dose calculations.

 \begin{figure}[!ht]\centering
\includegraphics[width=6cm]{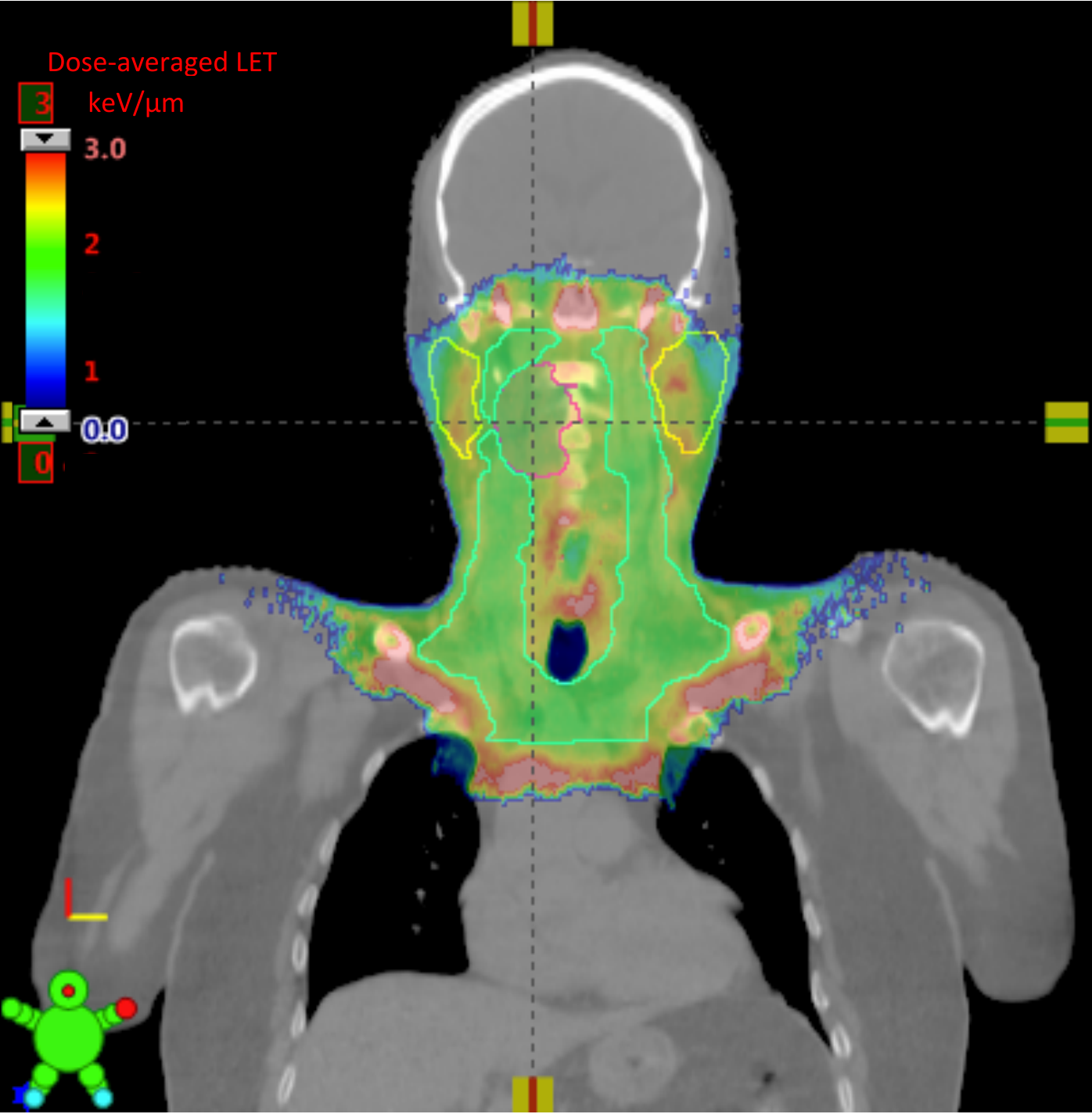}
\caption[]
{ Color wash plot showing our MC-calculated LET$_d$ map (in units of keV/$\mu$m), in the same coronal plane as in fig.~\ref{figure:barfigpaper}. }
  \label{figure:barlet}
 \end{figure}

As discussed above, one of the main strengths of this work is the ability to carry out detailed nuclear interaction simulations very rapidly on the GPU. It was shown in ref.\cite{Grassberger} that secondary protons have a significant impact on the LET$_d$ distribution in clinical proton beams. In fact, our MC is  capable of computing physical dose and LET$_d$ simultaneously; this does not considerably extend the net computational time. The LET$_{d}$ calculations in water are in good agreement with TOPAS. Fig.~\ref{figure:barlet} shows our prediction of the LET$_{d}$ distribution in areas with at least 10\% of the maximum dose, in the same coronal plane as in fig.~\ref{figure:barfigpaper}. As expected, LET$_{d}$ values are higher in regions where the protons are ranging out. At present, our LET$_{d}$ values neither include the small contributions of proton recoils from scattering of secondary neutrons, nor heavy ion recoils. The neutron scattering portion will be taken into account after the addition of a neutron transport kernel to our MC. The ability to perform MC-based LET$_{d}$ calculations rapidly and accurately on the GPU opens very exciting prospects for biological treatment planning.

%\subsection{Speed improvements}

\section{Conclusions}\label{section:conclusions}
To summarize, we have successfully developed a GPU-based proton transport MC that incorporates Bertini cascade and evaporation kernels for modeling non-elastic interactions of protons with any nucleus in the therapeutic energy range. We have verified our MC extensively using Geant4 and TOPAS. The net computation speed is typically $\sim$20 s per $1\times 10^7$ proton histories, which is comparable to processing times reported by previous authors using simulations with simpler nuclear interaction models. The very fast calculation speed allowed this MC to be used in our clinic as the central component of a treatment plan verification system, and also as the dose calculation engine for MC-based IMPT optimization. Furthermore, the detailed nuclear modeling not only gives us greater confidence in our physical dose calculations, but will allow us to perform accurate GPU-based MC calculations of LET$_{d}$, as well as fast neutron dose estimates.

\section{Acknowledgements}
This work was partially funded by a grant from Varian Medical Systems, Inc. We are grateful to our colleagues at Mayo Clinic for carefully reading the manuscript.

\end{document}